\documentclass[
twocolumn,%
showpacs,preprintnumbers,amsmath,amssymb,
superscriptaddress,unsortedaddress]{revtex4}

\usepackage{graphicx}
\usepackage{dcolumn}
\usepackage{bm}

\begin{document}


\title{Unmodulated spin chains as universal quantum wires}

\author{Antoni W\'ojcik}
\affiliation{Faculty of Physics,
 Adam Mickiewicz University,
Umultowska 85, 61-614 Pozna\'{n}, Poland.
}
\author{Tomasz {\L}uczak}
\altaffiliation{Corresponding author.
Phone: +48 (61) 829-5394,
fax: +48 (61) 829-5315.
E-mail: tomasz@amu.edu.pl}
\affiliation{Faculty of Mathematics and Computer Science,
Adam Mickiewicz University,
Umultowska 87, 61-614 Pozna\'{n}, Poland.
}
\author{Pawe{\l} Kurzy\'nski}
\affiliation{Faculty of Physics,
 Adam Mickiewicz University,
Umultowska 85, 61-614 Pozna\'{n}, Poland.
}
\author{Andrzej Grudka}%
\affiliation{Faculty of Physics,
 Adam Mickiewicz University,
Umultowska 85, 61-614 Pozna\'{n}, Poland.
}
\author{Tomasz Gdala}
\affiliation{Faculty of Mathematics and Computer Science,
 Adam Mickiewicz University,
Umultowska 87, 61-614 Pozna\'{n}, Poland.
}
\author{Ma{\l}gorzata Bednarska}
\affiliation{Faculty of Mathematics and Computer Science,
 Adam Mickiewicz University,
Umultowska 87, 61-614 Pozna\'{n}, Poland.
}

\date{May 11, 2005}%

\begin{abstract}
We study a quantum state transfer between two qubits interacting with the
ends of a quantum wire consisting of linearly arranged spins
coupled by an excitation conserving, time-independent Hamiltonian. We
show that if we control the coupling between the source and the destination 
qubits and the ends of the wire,
the evolution of the system can lead to an almost
perfect transfer even in the case in which all nearest-neighbour
couplings between the internal spins of the wire are equal.
\end{abstract}

\pacs{03.67.Hk, 03.67.Pp, 05.50.+q}
\maketitle

\newcommand{\ga}{\gamma}
\newcommand{\la}{\lambda}
\newcommand{\ran}{\rangle}
\newcommand{\lan}{\langle}
 
The problem of designing quantum networks which enable efficient
high-fidelity transfer of quantum states has recently been
addressed by a number of authors (see [1-23]). Ideally, such a
network should meet  both the simplicity and
the minimal control requirements. 
By simplicity we mean that the network
consists of typical elements coupled in a standard way so  that a
few networks can be combined together to form  more complex systems.
The minimal control requirement says that the transmission
of a quantum state through the network should be possible without
performing many control operations (as switching interactions on
and off, measuring, encoding and decoding, etc.). A 1D quantum
network (quantum wire) which fulfills both above requirements was
proposed by Bose [1] who considered a spin chain with the nearest
neighbour Heisenberg Hamiltonian; here the transmission of quantum
state between the ends of the chain was achieved simply by a free
evolution of the network. However, as was shown by Bose, if all
neighbour couplings have the same strength the fidelity of a
transmission decreases with the chain length $n$. A similar model
(with the Heisenberg Hamiltonian replaced by XY one) was
considered by Christandl {\em et al.} in \cite{CDEL}. They show
that one can transfer quantum states through arbitrary long chains
if  spin couplings are carefully chosen in a way depending on the
chain length $n$ (see also [3-8]). Note however that this
approach does not meet the simplicity requirement since one cannot
merge several ``modulated'' quantum wires into a longer one.

Here, we study a transfer of quantum states between two qubits
attached to the ends of a quantum wire consisting of $n$ linearly arranged spins.
In order to fulfill the 
requirement of simplicity we assume that all couplings  between
neighbouring spins forming the quantum wire are the same (and equal to $1$),
while the couplings  between the source and the destination qubits and the 
ends of the wire are equal to~$a$. We show that one can 
significantly improve of the fidelity of the transfer between the source
and the destination qubits by selecting the value of $a$ appropriately. 
In particular, choosing $a$ small enough,
one can achieve a transfer whose fidelity can be arbitrarily close to one, 
even for large $n$.

We assume that the Hamiltonian of the whole system of $n+2$
qubits conserves the number of excitation 
(e.g., it is a XY Hamiltonian), so the state
$\bigotimes_{j=0}^{n+1}|0\rangle_{j}$ is its eigenstate. 
Let $F_{ab}$ denote the fidelity of the transfer of
 of an arbitrary state
$|\phi\rangle=a|0\rangle+b|1\rangle$ from the source ($j=0$) to
destination ($j=n+1$) qubit and let $F=F_{01}$ stand for 
the fidelity of the state $|1\rangle$ transfer. It is easy to check
that the average fidelity $\langle F_{ab}\rangle$, where the average
is taken over all possible values of $a$ and $b$, 
is uniquely determined by~$F$, namely 
$$\langle F_{ab}\rangle=\frac{1}{3}+\frac{1}{6}{(1+F)^2}.$$ 
Thus, from now on, we shall consider only the evolution of the system 
with the initial state given by
$|\Psi(0)\rangle=|1\rangle_0|0\rangle_1\cdots |0\rangle_{n+1}$, i.e., 
the wire is in the polarized state.
Then, the state  space is spanned by vectors
$|p\rangle=\bigotimes_{j=0}^{n+1}|\delta_{jp}\rangle_{j}$,
$p=0,\dots,n+1$, and the Hamiltonian $H(a)$ of the system 
can be written as
$$H(a)=a(|0\ran \lan 1|+|n\ran\lan n+1|)+\sum_{p=1}^{n-1}|p\ran\lan p+1|+h.c.,$$
where, let us recall, both the source and the destination qubits are
coupled to the quantum wire with strength $a$, while all couplings
among spins of  the quantum wire are taken to be one. 
If $a^2\neq 2$, then all
the eigenvalues of $H(a)$ are of the form $\la=2\cos\ga$, where
$\ga$ is a solution of either of the two following equations
($\mu=\pm 1$):
$$\mu \cot (\ga) \cot^\mu
\Big(\frac{n+1}{2}\ga\Big)=\frac{a^2}{2-a^2}\,.$$
The eigenvector $|v^{(\gamma)}\ran$ corresponding to the
eigenvalue $2\cos \ga $ has  coordinates
$v_k^{(\gamma)}=\lan k| v^{(\gamma)} \ran$ given by
$$v_0^{(\gamma)}=\frac{a}{c}\sin \ga\,,$$
$$v_k^{(\gamma)}=\frac{1}{c}\Big(\sin \big((k+1)\ga\big)+(1-a^2)\sin\big((k-1)\ga\big)\Big)\,,$$
for $k=1,\dots,n$, and $$v_{n+1}^{(\gamma)}=\mu v_0^{(\gamma)} ,$$
where $c=c(a)$ is defined as
$$c^2=(n+1)(2(1-a^2)\cos^2\ga +a^4/2)+2 a^2-a^4.$$

\begin{figure}
{\includegraphics[width=8truecm]{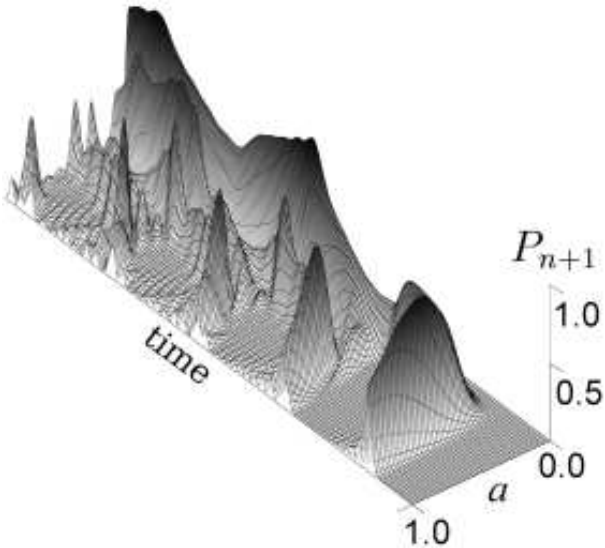}}
\caption{\label{f1} The evolution of $P_{n+1}(t)$ for the system of 30 spins for $t\in [0,150]$.}
\end{figure}

\begin{figure}
{\includegraphics[width=8truecm]{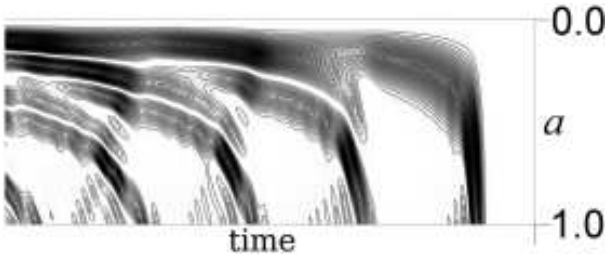}}
\caption{\label{f2} The density plot for the graph in Figure 1.}
\end{figure}

\begin{figure}
{\includegraphics[width=8truecm]{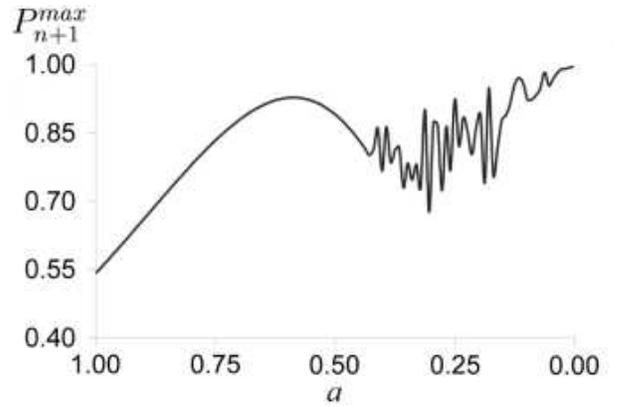}}
\caption{\label{f3} The dependence of $P_{n+1}^{\max}=\max\{P_{n+1}(t): t\in [0,20000]\}$ on $a$.
}
\end{figure}

Let $P_j(t)=|\lan j|\Psi(t) \ran|^2$ denote the probability 
of the excitation of the $j$th spin, if the initial state of the system 
is $|1\rangle_0|0\rangle_1\cdots |0\rangle_{n+1} $ (i.e., $P_0(0)=1$).
Figure~\ref{f1} shows the dependence of $P_{n+1}(t)$ on $a$
for the first period of the evolution of a system of 30 spins,  
Figure~\ref{f2} presents the density version of the 
same graph, while  Figure~\ref{f3} shows the projection of the evolution along the time axis.
As clearly seen in Figure~\ref{f3}, decreasing $a$ from $1$ to about $0.6$ 
significantly improves the fidelity of transfer; if we decrease $a$ even further,
then the fidelity drops down but for small $a$ it grows again approaching 
$1$ as $a\to 0$. 

\begin{figure}
{\includegraphics[width=8.2truecm]{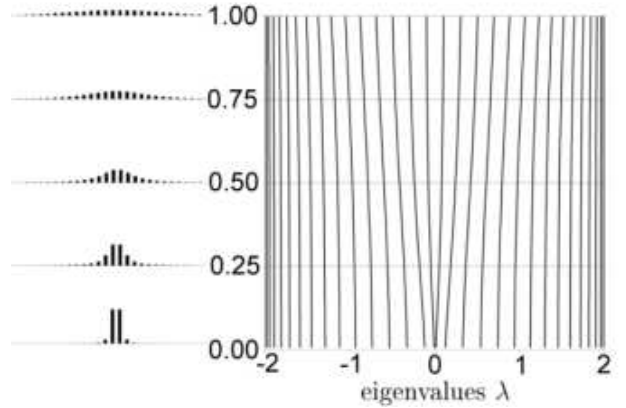}}
\caption{\label{f4} The dependence of the eigenvalues of $H(a)$ on $a$ (the right hand side
of the figure) and the eigenvector populations $\{|\langle v(\la)|\Psi(0)\rangle|^2\}_\lambda$ (on the left). }
\end{figure}

In order to better understand Figure~\ref{f3} we graph the effect
of decreasing $a$ on the eigenvalue spectrum of the whole system as
well as on the eigenvectors population $|v_0^{(\gamma)}|^2$
(Figure~\ref{f4}). For $a\sim 1$ the spectrum is nearly harmonic in
the vicinity of $0$ but the distribution of the eigenvectors is broad
and the contribution from the anharmonic part of the spectrum spoils
the transfer. If $a$ decreases the eigenvectors distribution
narrows so the anharmonic contribution drops down; consequently
the transfer fidelity increases. On the other hand, for very small
$a$, the harmonicity of the spectrum in the vicinity of zero
brakes down, which reduces the fidelity of the transfer.

For small values of  $a$  the behaviour of the system 
depends strongly on the parity of $n$. 
If $a \to 0+$, then the initial state $|\Psi(0)\ran$ is concentrated basically on 
two eigenvectors for $n$ even  (see Figure~\ref{f4}); if $n$ is odd, 
then the evolution of the system takes place in a three-dimensional space. 
More specifically, let  $\hat\la _a$ be the 
smallest positive eigenvalue of~$H(a)$. 
If $n$ is even and $a\to 0+$, then  for the two 
eigenvectors $|x \ran = |v(\hat\la _a) \ran$, 
$|y \ran = |v(-\hat\la _a) \ran$ corresponding to the eigenvalues $\pm \hat\la_a$, 
we have $x_0= y_0= x_{n+1}= -y_{n+1}\sim 1/\sqrt{2}$.
Consequently,  an almost perfect transfer of states between
qubits $0$ and $n+1$ occurs after time $\tau\sim \pi/(2\hat\la_a)$. 
For odd $n$'s, the spectrum of $|\Psi(0)\rangle$ is concentrated on three 
eigenvectors: besides $|x \ran= |v(\hat\la _a) \ran$ and $|y \ran = |v(- \hat\la _a) \ran$, 
we need to take into account the eigenvector $|z\ran = |v(0) \ran$
corresponding to the eigenvalue zero. Then, $z_0= |z_{n+1}|\sim  1/\sqrt{2}$,
$x_0= y_0=|x_{n+1}|= |y_{n+1}|\sim 1/2$, and the sign of $z_{n+1}$ is
opposite to the signs of both $x_{n+1}$ and $y_{n+1}$. In this case
an almost perfect transfer from the source to the destination qubits occurs
after time $\tau\sim\pi/\hat\la_a$. Let us mention that a similar phenomenon 
of transferring the states between two qubits weakly interacting with a network
has been recently observed by Li {\sl et al.}~\cite{Li} and 
Plenio and Semiao \cite{Plenio}, who  consider transferring qubit 
states through a ladder and a cycle, respectively.

\begin{figure}
{\includegraphics[width=8.2truecm]{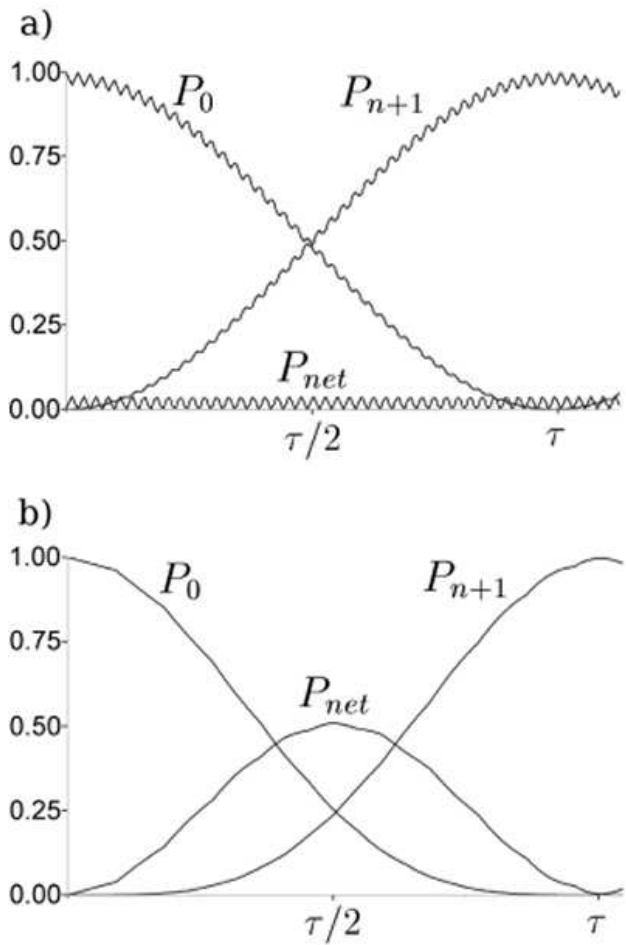}}
\caption{\label{f5} The evolution of  $P_0(t)$, $P_{n+1}(t)$, $P_{net}(t)$, for $a=0.01$ 
and (a)  $n=198$; (b) $n=199$. If $n=198$, then $\lambda_a\sim 10^{-4} $ and $\tau\sim 16000$; 
for $n=201$ we have $\la_a =1.41\cdot 10^{-3} $ and $\tau\sim 2200 $.}
\end{figure}

Straightforward computations show that the fidelity of the
transfer scales as $1-O(a^2 n)$ and it grows to one only when
$a\ll1/\sqrt{n}$. Moreover,  we have
$\hat\la_a\sim a^2 $ for an even $n$, and $\hat \la_a\sim
2a/\sqrt{n}$ when $n$ is odd. Hence, the time of the transfer is $\Theta(a^{-2})$
for even $n$, and $\Theta(\sqrt n/a)$ for odd $n$. Thus,  for
$a=\delta / \sqrt{n}$, an almost perfect
($F\approx1-\delta^2$) transfer is possible for $\delta\ll1$ in a
time which scales linearly with the quantum wire length $n$. The
speed of the transfer $n/\tau$ is approximated by $2
\delta/\pi$ and $\delta^2 /\pi$ when $n$ is  odd and even, respectively.
Thus, for small $a$,  a quantum wire with odd $n$ ensures more effective transfer.
In Figure~\ref{f5} the evolution of the excitation probabilities  of the source
qubit $P_0(t)$, the destination qubit $P_{n+1}(t)$ and the total probability 
of the excitation of the wire qubits  $P_{net}(t)=\sum_{j=1}^n P_j(t)$,
are presented for $a=0.01$ and  $n=198,199$. Note that if $n$ is even,
the spins of the wire remain almost unexcited during the
evolution, i.e., the source and destination qubits evolve like two
isolated and directly coupled qubits. Thus, the evolution of a system 
reminds  Rabi oscillations between the source and destination qubits, 
despite the fact that they lie in a significant distance from each other. 

Finally, we remark   that if $n$ is even and $a\to 0+$, after time $t\sim \tau/2$
the state of the system is close to $\tfrac{1}{\sqrt{2}}(|0 \ran_0|1 \ran_{n+1}+
|1 \ran_0|0\ran_{n+1})|0\ran_1\cdots|0\ran_n$. 
Thus, such an unmodulated quantum wire  can also be used 
to generate entanglement between
the source and destination qubits. 

In conclusion, we have shown that one can significantly improve a
transfer of the qubit states  between the ends of a quantum wire by
controlling the coupling on its ends. In particular, an unmodulated
spin chain of arbitrary length $n$ can be used as a universal
quantum wire to transfer of quantum states with arbitrary
high fidelity in a time which scales linearly with $n$.

\smallskip

The authors  wish to thank the State Committee for Scientific Research
(KBN) for its support: 
 A.W.\ and A.G.\ were supported by grant 0~T00A~003~23;
T.\L., T.G.,\ and M.B. by grant 1 P03A 025 27.

\end{document}